\documentclass[nofootinbib,aps,prl,twocolumn,amsmath,amssymb,floatfix,superscriptaddress,]{revtex4-1}

\usepackage[utf8]{inputenc}
\usepackage[dvipsnames]{xcolor}
\usepackage{graphicx}
\usepackage{amsmath}
\usepackage{amsthm}
\usepackage{dsfont}
\usepackage{bbold}
\usepackage{braket}
\usepackage{bm}
\usepackage{hyperref}

\newtheoremstyle{mystyle}%
{3pt}%
{3pt}%
{\upshape}%
{}%
{\bfseries}%
{.}%
{.5em}%
{}%

\usepackage[all]{hypcap}
\usepackage[capitalise,compress]{cleveref}
\crefname{section}{Section}{Section} %
\crefname{subsection}{Section}{Section}
\theoremstyle{mystyle}
\crefname{thm}{Theorem}{Theorems}
\Crefname{thm}{Theorem}{Theorems}

\crefname{corll}{Corollary}{Corollaries}
\theoremstyle{remark}

\theoremstyle{definition}

\theoremstyle{mystyle}

\DeclareMathOperator{\tr}{tr}

\begin{document}

\title{Quantum Sensing with Erasure Qubits}

\author{Pradeep Niroula}
\affiliation{Joint Center for Quantum Information and Computer Science, NIST/University of Maryland, College Park, Maryland 20742, USA}
\affiliation{Joint Quantum Institute, NIST/University of Maryland, College Park, Maryland 20742, USA}
\author{Jack Dolde}
\affiliation{Department of Physics, University of Wisconsin-Madison, Madison, Wisconsin 53706, USA}
\author{Xin Zheng}
\affiliation{Department of Physics, University of Wisconsin-Madison, Madison, Wisconsin 53706, USA}
\author{Jacob Bringewatt}
\affiliation{Joint Center for Quantum Information and Computer Science, NIST/University of Maryland, College Park, Maryland 20742, USA}
\affiliation{Joint Quantum Institute, NIST/University of Maryland, College Park, Maryland 20742, USA}
\author{Adam Ehrenberg}
\affiliation{Joint Center for Quantum Information and Computer Science, NIST/University of Maryland, College Park, Maryland 20742, USA}
\affiliation{Joint Quantum Institute, NIST/University of Maryland, College Park, Maryland 20742, USA}
\author{Kevin C. Cox}
\affiliation{
DEVCOM Army Research Laboratory, Adelphi, Maryland 20783, USA}
\author{Jeff Thompson}
\affiliation{Department of Electrical and Computer Engineering, Princeton University, Princeton, NJ, 08544, USA}
\author{Michael J. Gullans}
\affiliation{Joint Center for Quantum Information and Computer Science, NIST/University of Maryland, College Park, Maryland 20742, USA}
\author{Shimon Kolkowitz}
\affiliation{Department of Physics, University of Wisconsin-Madison, Madison, Wisconsin 53706, USA}
\author{Alexey V. Gorshkov}
\affiliation{Joint Center for Quantum Information and Computer Science, NIST/University of Maryland, College Park, Maryland 20742, USA}
\affiliation{Joint Quantum Institute, NIST/University of Maryland, College Park, Maryland 20742, USA}

\date{\today}

\begin{abstract}
The dominant noise in an ``erasure qubit'' is an erasure---a type of error whose occurrence and location can be detected. Erasure qubits have potential to reduce the overhead associated with fault tolerance. To date, research on erasure qubits has primarily focused on quantum computing and quantum networking applications. Here, we consider the applicability of erasure qubits to quantum sensing and metrology.
We show theoretically that, for the same level of noise, an erasure qubit acts as a more precise sensor or clock compared to its non-erasure counterpart. 
We experimentally demonstrate this by artificially injecting either erasure errors (in the form of atom loss) or dephasing errors into a differential optical lattice clock comparison, and observe enhanced precision in the case of erasure errors for the same injected error rate. 
Similar benefits of erasure qubits to sensing can be realized in other quantum platforms like Rydberg atoms and superconducting qubits.
\end{abstract}

\maketitle

Noise, i.e.,~environment-induced decoherence, presents a fundamental challenge in quantum sensing. While noiseless sensing can exhibit so-called Heisenberg scaling in precision when using appropriately optimized probe states~\cite{bollinger96,wineland92,giovannetti2004quantum,giovannetti2006quantum}, noise typically leads to a worse scaling~\cite{escher2011general, demkowicz2012elusive,kolodynski2013efficient}. Given certain assumptions on noise, we can regain Heisenberg scaling using an appropriate error correcting code~\cite{demkowicz2017adaptive,zhou2018achieving,layden2019ancilla}, but such schemes require a costly overhead in the ancilla and/or the operations necessary for error detection and correction. This places practical limits on how well such schemes can improve metrological performance~\cite{shettell2021practical}; furthermore, finding a noise-appropriate error-correction code can itself be a challenge. 

A complementary approach to noise resilience is to engineer ``erasure qubits'' where the dominant noise is an erasure error---a type of error that takes the qubit out of the computational space and whose occurrence and location can be detected \cite{grassl97}. A simple example of an erasure qubit is a photon encoded in the polarization basis (i.e., H/V), where the absence of a photon can be used to detect photon loss \cite{bennett2014quantum, knill01}. Recent work has extended this concept to design qubit encodings in other platforms that result in the conversion of the dominant errors into erasures \cite{wu2022erasure}. This has been proposed and demonstrated for neutral-atom \cite{wu2022erasure,sahay23,ma23,scholl23} and superconducting \cite{kubica2022erasure,levine2023demonstrating,teoh2022dual} qubits, and also proposed for trapped ions \cite{kang23a}.

\begin{figure}
    \centering
    \includegraphics[width=0.48\textwidth]{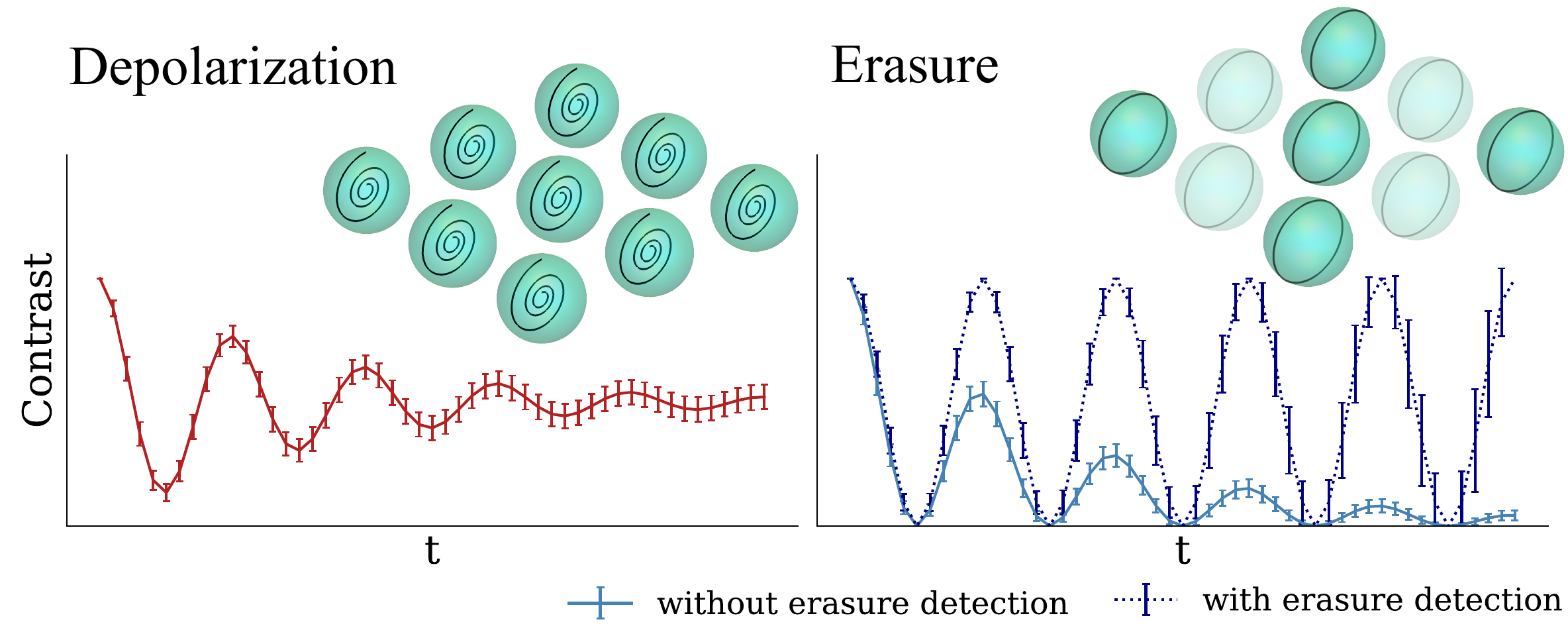}
    \caption{Illustration of Ramsey interrogation of an ensemble of two-level sensors undergoing depolarizing (left) and erasure (right) noise. Time evolution of the quantum sensors is represented by black lines inside the Bloch spheres. Erasure errors are shaded out in the right ensemble. Bottom plots show respective fringe contrasts.}
    \label{fig:schematic}
\end{figure}

Erasures are easier to protect against than errors with unknown location. An error correcting code of distance $d$ can correct only $\lfloor (d-1)/2 \rfloor$ errors with unknown locations, but can correct $d-1$ erasures \cite{gottesman1997stabilizer}. Consequently, the two-qubit gate error-rate threshold is higher for erasure qubits than for general qubits, sometimes allowing a lower overhead in implementing correction schemes.

In this work, we show that, similarly to the case of a quantum computer, 
not all noise processes degrade the performance of a quantum sensor in the same way and that using erasure qubits can improve sensor performance. %
The corresponding performance gain can be quantified as an increase in  Fisher information. In particular, we show that the uncertainty bounds for noisy sensing, given by the single-parameter Cram\'{e}r-Rao bound, can be made tighter with erasure qubits.  Importantly, while fault-tolerant error correction demands continuous monitoring of the system to control the growth of errors, in quantum sensing, it is often sufficient to isolate the erroneous data, preventing it from contributing to the signal used for parameter estimation. 
We also use an optical lattice clock to confirm experimentally that erasure errors have a fundamentally different effect on sensing precision than other sources of decoherence (see Fig.~\ref{fig:schematic} for illustration).

\medskip 

\noindent \emph{Noisy Sensing.---}While our arguments extend to general, multi-qubit sensing, we focus on a single-qubit sensor coupled to an unknown parameter $\phi\in[0,2\pi)$ via the generator $\sigma_z/2$ and subject to a noise process. An input sensor state evolves under the unitary  $\exp(-i\phi \sigma_z/2)$ but can undergo a noise process $\mathcal{E}$ with a certain probability, $q$, leading to the channel: $\rho_0 \to (1-q)e^{-i\phi \sigma_z/2}\rho_0  e^{i\phi \sigma_z/2} + q\mathcal{E}_{\phi}(\rho_0)$.

Measurements are performed on the final state to obtain an estimate $\hat{\phi}$. For noiseless sensing, the optimal sensing protocol involves initializing the sensor as $\ket{+} = (\ket{0}+\ket{1})/\sqrt{2}$, letting it accumulate relative phase, and measuring it in the $\ket{\pm}$ basis~\cite{giovannetti2004quantum}. We note that, when multiple entangled probes are used or the encoding unitary is applied many times, a ``phase wrapping'' issue arises where one can lose track of which $2\pi$ interval the relative phase is in~\cite{higgins2009demonstrating, hayashi2018resolving}, requiring more involved measurement schemes ~\cite{kimmel2015robust, kimmel2015robusterratum, belliardo2020achieving}.

Assuming that the sensing is unbiased, that is $\mathbb{E}[\hat{\phi}]=\phi$, the uncertainty in $\hat{\phi}$ is lower-bounded by the quantum Cram\'{e}r-Rao bound: $(\delta\phi)^2 = \mathbb{E} (\phi - \hat{\phi})^2 \geq (\mu\mathcal{F}(\phi;\rho))^{-1}$, where $\mathcal{F}(\phi;\rho)$ is the  quantum Fisher information and $\mu$ is the number of measurements.
This single-parameter estimation bound is known to be saturable asymptotically in $\mu$. In this work, we focus on the limit where the number of measurements is large enough for asymptotic arguments to hold. 

The single-parameter quantum Fisher information is a convex quantity; that is, for a density matrix $\rho = \alpha \rho_1 + \beta\rho_2$, we have $\mathcal{F}(\alpha\rho_1 + \beta \rho_2) \leq \alpha \mathcal{F}(\rho_1) + \beta\mathcal{F}(\rho_2)$. If we assume that the error state $\mathcal{E}_{\phi}(\rho_0)$ does not carry any information about $\phi$, i.e., $\mathcal{F}(\mathcal{E}_{\phi}(\rho_0))= 0$, we get an upper-bound on the noisy sensor $\mathcal{F}((1-q)\rho_{\phi} + q\mathcal{E}_{\phi}(\rho_0)) \leq (1-q) \mathcal{F}(\rho_\phi)$. However, as we show below, while this bound is not attainable using the straightforward sensing scheme mentioned above, it is attainable using sensors based on erasure qubits. %

\medskip 

\noindent \emph{Depolarizing and Dephasing Noise.---}Consider the simplest form of noise where random Pauli operators act on the sensor with equal probability. This gives rise to a depolarizing noise of strength $q$ with $\mathcal{E}_{\phi}(\rho_0) = \mathbb{1}/2$. In the Supplemental Material, we show that the optimal input state for sensing under depolarizing noise is the same as for sensing without noise, that is, $\rho_0 = \ket{+}\bra{+}$. For generality, we measure the final state in the basis given by $\ket{\pm \theta} = (\ket{0} \pm e^{i\theta}\ket{1})/\sqrt{2}$. The two outcomes have probabilities  $p_{\pm} =  (1 \pm (1-q) \cos (\phi-\theta))/2$ and the corresponding Fisher information is given by
\begin{align}
    \mathcal{F} &= \left\langle \left({\partial_{\phi} \log p_x }\right)^2\right\rangle_{x = \pm}= \frac{(1-q)^2\sin^2{(\phi-\theta)}}{1-(1-q)^2\cos^2{(\phi-\theta)}}.
    \label{eq:fisher-depol}
\end{align}
Choosing a $\phi$-dependent basis achieves the maximum Fisher information of $(1-q)^2$ \cite{kolodynski2013efficient}. It is, however, not always possible to have prior information about $\phi$. Therefore, the achievable Fisher information is usually lower than the maximal value of $(1-q)^2$. 

If we instead consider a purely dephasing channel, with $\sigma_z$ operator acting with probability $q$, the Fisher information is still Eq.~(\ref{eq:fisher-depol}), but with a modified strength $2q$.
Dephasing and depolarizing noise are often used to model realistic quantum devices \cite{krantz2019quantum, nielsen2010quantum}. In both of these cases,  the Fisher information scales quadratically in $(1-q)$, meaning we fail to saturate the bound dictated by the convexity \cite{kolodynski2013efficient}.

\medskip 

\noindent \emph{Erasure.---}Now, consider a noise process that takes the sensor to a third state $\ket{-1}\bra{-1}$ that can be detected using non-demolition measurements without perturbing the coherence between the computational states $\ket{0}$ and $\ket{1}$  used in sensing, i.e, $\mathcal{E}_{\phi}(\rho_0) = \ket{-1}\bra{-1}$. When the sensor is equipped with erasure conversion, we add an erasure detection step to the usual measurement protocol; if we detect the erasure state $\ket{-1}\bra{-1}$, we do nothing (record ``null''). Otherwise, we measure in the $\ket{\pm}$ basis as before. While in quantum computing applications, it is important to have mid-circuit measurement of erasure errors with no back-action onto the
qubit levels \cite{ma23, levine2023demonstrating}, the measurements in the quantum sensing protocols considered here are terminal measurements. Therefore, it suffices to merely distinguish $\ket{\pm}$ from erasure in the final measurement.

The erasure detection step prevents experimental errors from creeping into the measurements used to derive the estimator $\hat \phi$. The three outcomes of this sensing protocol are $\ket{-1}, \ket{+}$ and $\ket{-}$, with measurement probabilities $p_{-1} = q$ and $p_{\pm} = (1-q)(1 \pm \cos (\phi))/2$, and the Fisher information associated with these three outcomes is $\mathcal{F}_{\rm erasure} = (1-q)$, attaining the maximal linear scaling in $(1-q)$. 

\medskip 

\noindent \emph{Erasure Errors in Atomic Clocks.---}Today, atomic clocks can achieve a precision corresponding to an uncertainty of less than 1 second over the lifetime of the universe \cite{ludlow2015optical, bloom2014optical, bothwell2019jila}. This precision has enabled tests of foundational physical theories such as special and general relativity~\cite{bothwell2022resolving, ZhengRedshift, hafele1972around, takamoto2020test}. Continuing advances in atomic clocks' performance promise to make them prime platforms for tests of fundamental physics. 

Atomic clocks work by measuring the deviation of the frequency of a local oscillator $f_{LO}$ from a narrow transition line. In an optical atomic clock, a laser serves as the local oscillator, and is used to drive a stable transition with reference frequency $f_0$ .
The laser frequency is stabilized by measuring the shift $\Delta f = f_{LO}-f_0$ with respect to the clock frequency using Ramsey spectroscopy. After each Ramsey interrogation, the clock applies an electronic correction to the laser source to compensate for frequency drifts. 

Measuring a frequency shift using Ramsey spectroscopy amounts to phase estimation and has the same structure of the sensing problem outlined above. The relevant figure of merit for an atomic clock is the ``fractional instability'' $ \sigma = \delta f_{LO}/{f_0}$, where $\delta f_{LO}$ is the uncertainty in the measurement of the local oscillator frequency. 

Historically, the performance of optical atomic clocks has been limited not by the reference atoms, but rather by noise from the laser probe. However, recent experiments used correlated differential spectroscopy to bypass the limitations of the local oscillator and achieve differential clock comparisons that are limited by Quantum Projection Noise (QPN) of the atoms \cite{zheng2022differential, bothwell2022resolving}. This provides an opportunity to explore and reduce instability arising from errors acting on the atoms. 

A Ramsey-like protocol is used to measure the phase $\phi$ accumulated over the measurement time $T_c$. $\phi$ is related to the frequency difference between the oscillator and the atom as $\phi = 2\pi T_c (\Delta f)$; the local oscillator frequency is thus inferred as $f_{LO} = f_0 + \phi/(2\pi T_c)$. The fractional instability is then expressed in terms of the uncertainty in the estimate of $\phi$, which can, in turn, be bounded by Fisher information using the Cram\'{e}r-Rao bound,
\begin{equation}
     \sigma = \frac{\delta f_{LO}}{f_0} = \frac{1}{2\pi T_c}\frac{\delta \phi}{f_0}, \qquad  %
     \delta \phi \geq \frac{1}{\mathcal{F}^{1/2}},
     \label{eq:stability-definition}
\end{equation}
where $\mathcal{F}$ is the single-parameter Fisher information associated with measurement of the phase shift $\phi$. Ramsey interrogation of the clock state is similar to the single-parameter sensing problem of calculating the unknown parameter $\lambda$ in the Hamiltonian $H=\lambda\sigma_z/2$ driving a two-level system. The optimal measurement protocol, in both cases, is to start in state $\ket{+}$, evolve freely under the Hamiltonian, %
and measure the resulting state in the $\ket{\pm}$ basis. %
The quantum information associated with a single measurement of the parameter is $\mathcal{F} = 1$ (assuming no noise).

In an optical lattice clock, on each interrogation, we measure the ensemble of $N$ atoms, and we repeat this $\tau/T_c$ times, where $\tau$ is the total measurement time and $T_c$ is the time of each measurement cycle. For independent sensors, Fisher information increases additively, giving a lower bound on the fractional instability: 
\begin{equation}
   \sigma  = \frac{1}{2\pi T_c}\frac{\delta \phi}{f_0} \geq \frac{1}{2\pi}\frac{1}{f_0}\sqrt{\frac{1}{N}\frac{1}{T_c \tau}}.
   \label{eq:stability-noiseless}
\end{equation}

We can now consider how the bounds on fractional instability change in the case of i) noise that keeps the sensor state within the sensing subspace and cannot be detected, and ii) erasure noise that takes the quantum state out of the sensing subspace. 

For the former case, consider a dephasing noise model that decoheres a state at a fixed rate. Dephasing noise in an atomic clock can arise from inhomogeneous light shifts from the lattice, line-broadening from atomic collisions, or a magnetic field gradient \cite{zheng2022differential}. During Ramsey spectroscopy, dephasing noise has the same effect as a fully depolarizing channel [see discussion following Eq.\ (\ref{eq:fisher-depol})]; after interrogation time $T_c$, the clock state is assumed to be in a depolarized state $\mathbb{1}/2$ with probability $q = 1-e^{-\Gamma T_c}$. With probability $1-q$, the clock state stays intact. Succinctly, the noise channel is modeled as $\rho(t) = (1-q) \rho_{\phi}+ q{\mathbb{1}}/{2}$, with $q = 1-e^{-\Gamma T_c}$.

The error rate $q$ is obtained by measuring the Ramsey fringe contrast $C$, which decays as $ \propto e^{-\Gamma_c t}$. Using this rate in Eq.~\eqref{eq:fisher-depol},  the Fisher information of this noisy sensor is $\mathcal{F}_{\rm depol} \leq (1-q)^2$, giving the lower bound on fractional instability
\begin{equation}
   \sigma_{\rm depol}  \geq \frac{1}{1-q}\frac{1}{2\pi}\frac{1}{f_0}\sqrt{\frac{1}{N}\frac{1}{T_c \tau}} = \frac{\sigma}{1-q},
   \label{eq:stability-depol}
\end{equation}
where the depolarization strength $q$ is determined by the interrogation time $T_c$. 

For erasure noise models, similar to dual-rail photonic qubits for which photon loss is an erasure error, we can consider errors due to loss of the atoms from the lattice during the measurement, or equivalently due to imperfect initialization of the atoms at the start of the measurement. Assume that each atom is lost during clock interrogation (or is incorrectly initialized) with probability $q$. In contrast to a bit-flip or a depolarizing noise, such a noise takes the sensor out of the computational subspace, and can be accounted for in subsequent data processing. Consequently, lost or incorrectly initialized atoms do not contribute to the signal used to estimate the accumulated phase, leading to a better signal-to-noise ratio. The Fisher information in this case is $\mathcal{F} = (1-q)$, giving a lower bound 
\begin{equation}
   \sigma_{\rm atom-loss}   \geq \frac{1}{\sqrt{1-q}}\frac{1}{2\pi}\frac{1}{f_0}\sqrt{\frac{1}{N}\frac{1}{T_c\tau}} = \frac{\sigma}{\sqrt{1-q}}.
   \label{eq:stability-erasure}
\end{equation}
The two noise processes therefore contribute differently to the fractional instability. 

\begin{figure*}
    \centering
\includegraphics[width=0.9\textwidth]{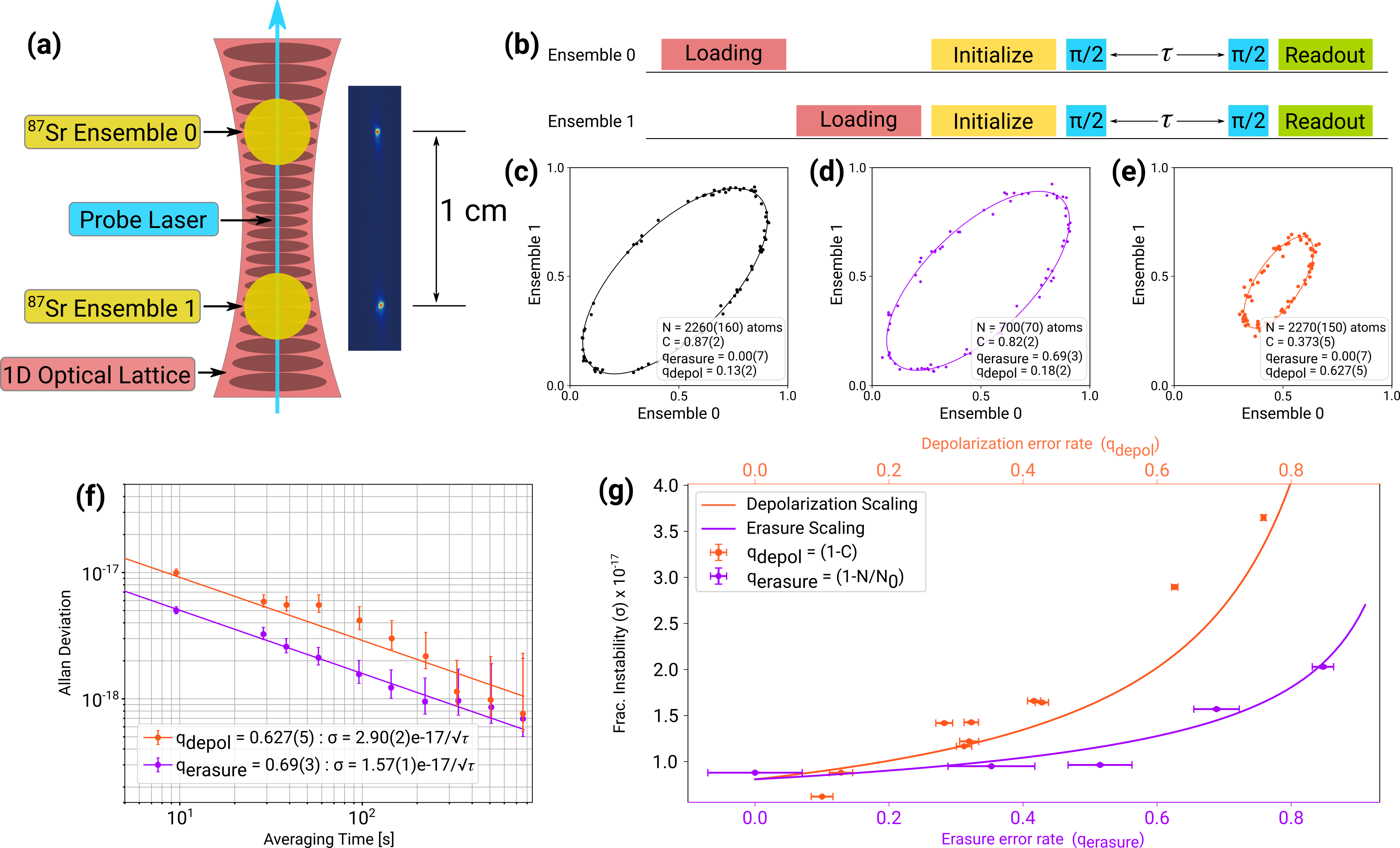}
    \caption{(a) Diagram of two ensembles of $^{87}$Sr loaded into the same $1$-D optical lattice trap. The ensembles, separated vertically by $1$ cm, can be interrogated simultaneously with a laser directed along the axis of the trap. A camera image of the ensembles is shown on the right. (b) Experimental sequence (timing not to scale) used for differential comparisons of the ensembles. The atoms are initialized in the ground state, then synchronously probed with a Ramsey interrogation technique. We choose a Ramsey dark time, $\tau$, of 8.0 seconds. The populations in the ground and excited states are then measured. The total sequence time is 9.67 seconds. (c) The excitation fraction of each ensemble is plotted parametrically, tracing out an ellipse corresponding to the differential phase between the ensembles. (d) The atom number, $N$, in this experiment is decreased relative to the ellipse in (c) while keeping the contrast, $C$, constant. This ellipse corresponds to an increase in erasure error. (e) The contrast, $C$, is decreased in this experiment relative to the ellipse in (c), while keeping atom number, $N$, approximately the same. This ellipse corresponds to an increase in depolarization error. (f) Extracted stabilities from ellipses shown in (d) and (e). Even with similar error rates, the erasure error leads to lower instability than depolarization error. (g) Measured fractional instability against error rate $q$ for erasure error and depolarization error. For the depolarizing noise, the error rate $q$ is derived from the differential Ramsey fringe contrast as $q = 1-C$. For atom loss, the error rate is taken to be $q = 1-N/N_0$, where $N_0$ is the average number of atoms initialized in both ensembles, and $N$ is the average number of atoms remaining after each Ramsey interrogation. The solid lines are obtained using a least-square fit to both data sets with the expected scaling of Fisher information given by Eqs.~\eqref{eq:stability-erasure} and \eqref{eq:stability-depol}, respectively, with the fractional instability at $q=0$ as the only free parameter.
}
    \label{fig:erasure-clocks}
\end{figure*}

We study the response of fractional stability to error rates for the two types of errors using a differential clock comparison experiment in a multiplexed optical lattice clock \cite{bothwell2022resolving,zheng2022differential} as illustrated in Fig.~\ref{fig:erasure-clocks}(a). Two spatially-resolved ensembles of ${}^{87}$Sr atoms are used, and their \textit{relative} frequency shift is measured through synchronous Ramsey interrogation with the same clock laser as shown in Fig.~\ref{fig:erasure-clocks}(b). Details regarding this technique and the experimental apparatus are discussed in Ref.~\cite{zheng2022differential}.  The differential frequency between the ensembles can be determined by parametrically plotting the measured excitation fractions from each experiment and fitting to the resulting ellipse, as shown in Fig.~\ref{fig:erasure-clocks}(c). The synchronous measurement overcomes the limitation placed on interrogation time by the frequency instabilities (line-width) of the laser. Denoting the frequency shifts of the ensembles by $\phi_a$ and $\phi_b$, the fractional instability in Eq.~\eqref{eq:stability-definition} thus becomes $\delta(\phi_a-\phi_b)$. Assuming the ensembles have the same number of atoms and the same coherence, the variance in the relative shift is the sum of individual variances, $\text{Var}(\phi_a-\phi_b) = \text{Var}\left(\phi_a\right) + \text{Var}\left( \phi_b\right) = 2\text{Var}\left(\phi_b\right)$. This contributes a factor of $\sqrt{2}$ to Eq.~\eqref{eq:stability-noiseless}. The scaling of instability with noise in Eqs.~\eqref{eq:stability-depol} and \eqref{eq:stability-erasure} remains unchanged. 

In one set of experiments, an example of which is shown in Fig.~\ref{fig:erasure-clocks}(d), we tune the erasure error rate while keeping the coherence time of the atoms fixed. In order to controllably vary the erasure loss rate, we can intentionally introduce an error in our atom initialization, which is equivalent to a noise model where atoms are lost during the experiment.  Any atoms not initialized in the correct hyperfine ground state are removed from the optical lattice and do not contribute to the estimate of the accumulated phase difference. On average $N_0$ atoms are loaded in each experiment before initialization. The average number of atoms participating in phase estimation is used to derive the probability of an erasure error, i.e, $q = \langle N \rangle/\langle N_0 \rangle$. 

In another set of experiments, an example of which is shown in Fig.~\ref{fig:erasure-clocks}(e), we intentionally induce decoherence of the atomic superposition while holding the erasure rate steady. This is achieved by detuning the wavelength of the optical lattice away from the ``magic wavelength" \cite{ludlow2015optical} to induce an inhomogenous lattice light shift as the atoms experience different lattice trap depths due to their finite radial temperature, resulting in a dephasing of the ensembles. The strength of effective decoherence is then measured using Ramsey fringe contrast, i.e., $q = 1-C$. 

In Fig.~\ref{fig:erasure-clocks}(f), we plot the Allan deviation, experimentally calculated using a jackknifing technique \cite{young2020half, marti2018imaging}, from the ellipses in Figs.~\ref{fig:erasure-clocks}(d) and (e). In Fig.~\ref{fig:erasure-clocks}(g), we plot the extracted differential instability of the clock comparison against the error rates for dephasing and erasure errors. We observe that the measured instabilities are consistent with the scalings predicted by the lower bounds on Fisher information given by Eqs.~\eqref{eq:stability-depol} and \eqref{eq:stability-erasure}.

Several recent optical lattice clock experiments have already implicitly taken advantage of erasure errors by demonstrating atom-atom coherence times significantly exceeding the lifetime of atoms~\cite{bothwell2022resolving,zheng2022differential}. Future experiments could benefit by engineering erasure conversion of errors due to lattice Raman scattering and spontaneous emission. While Fig.~\ref{fig:erasure-clocks}(g) appears to promise dramatic improvements in fractional instability by converting depolarizing errors into erasure errors for high error rates $q>0.5$, it is important to note that this high error regime is not where clocks typically operate, and furthermore that the bounds shown in the figure and given by Eqs.~\eqref{eq:stability-depol} and \eqref{eq:stability-erasure} are for a fixed interrogation time $T_c$. In reality, $T_c$ should always be adjusted to minimize the clock instability for a given type of error and error rate. For a  QPN-limited, zero-dead-time differential clock comparison subject to depolarization errors with an exponential decay in contrast $C$ with rate $\Gamma_d$, $C=C_o e^{-T_c\Gamma_d}$, it can be shown that converting all depolarizing errors into erasure errors and re-optimizing the interrogation time results in at most a factor of $\sqrt{2}$ reduction in instability. For finite dead times, the improvement in instability can be larger, as erasure errors enable longer interrogation times without significant degradation in instability. For a QPN-limited clock comparison in the limit of long dead times $T_d \gg 1/\Gamma_d$, the reduction in instability from complete erasure conversion for optimized coherent interrogation times asymptotically approaches $2$.

\medskip

\medskip 

\noindent \emph{Discussion.---}In this work, we discuss erasure errors in the context of quantum sensing. We relate the metrological gain for erasure qubits with the saturation of Fisher information. We also discuss erasure errors in %
atomic clocks %
and %
experimentally demonstrate the different ways by which general errors and erasure errors affect the clock stability.

Similar benefit may be realized in sensing with other multi-level quantum systems. An erasure qubit composed of a ${}^3 P_0$ state and a Rydberg state in an alkaline-earth(-like) atom \cite{wu2022erasure,sahay23,ma23,scholl23} can be used to measure electric fields.
A dual-rail superconducting erasure qubit  \cite{kubica2022erasure,levine2023demonstrating, teoh2022dual} %
can be used to measure the coupling strength between the two constituent qubits. Finally, an erasure qubit based on the ground and the second excited state of a transmon \cite{kubica2022erasure} can be used to measure the transmon frequency (and potentially a magnetic field oscillating at a frequency larger than the frequency that dominates qubit dephasing) or a two-photon Rabi frequency coupling the two states.

Additionally, quantum sensors may be used as spectator qubits, where they are embedded into a quantum computer among `data qubits', which perform the actual quantum computation, to sense noise, drifts, and fluctuations \cite{majumder2020real}; this allows for feedback-based error-mitigation. Improving the performance of such sensors, by engineering erasure conversion, is a promising approach towards near-term error resilience.

\begin{acknowledgements}

P.N., J.B., A.E., and A.V.G.~were supported in part by AFOSR MURI, AFOSR, DARPA SAVaNT ADVENT, NSF PFCQC program, ARO MURI, DoE ASCR Accelerated Research in Quantum Computing program (award No.~DE-SC0020312), NSF QLCI (award No.~OMA-2120757), U.S.~Department of Energy Award No.~DE-SC0019449, and the DoE ASCR Quantum Testbed Pathfinder program (award No.~DE-SC00119040). Support is also acknowledged from the U.S.~Department of Energy, Office of Science, National Quantum Information Science Research Centers, Quantum Systems Accelerator. J.T.~acknowledges support by NSF QLCI (award No.~OMA-2120757). J.D.,  X.Z., and S.K.~were supported by a Packard Fellowship for Science and Engineering, the Sloan Foundation, the Army Research Office through agreement number W911NF-21-1-0012, and the National Science Foundation under Grants No.~2143870 and 2326810.

\end{acknowledgements}

\bibliography{citations}

\section{Supplemental Material}

\subsection{Optimal Input States for Noisy Sensing}
\label{sec:noisy-optimal-state}
In this section, we show that the optimal input state for noisy sensing of a single parameter under depolarizing noise is the same as that for noiseless sensing. 

Consider a single-qubit input state, $\rho_0$, which evolves under the sensing unitary $U_{\phi} = \exp(i\phi H)$, with $\phi$ being the sensing parameter and $H$ being the generator Hamiltonian. The quantum Fisher information, as a function of the input state $\rho_0$, with respect to the parameter $\phi$, is given by \cite{liu2020quantum}
\begin{equation}
    \mathcal{F}(\rho_0) = 4(2\tr(\rho_{0}^2)-1)|\braket{\eta_0|H|\eta_1}|^2,
    \label{eq:qfi}
\end{equation}
where $\eta_0$ and $\eta_1$ are the eigenvectors of $\rho_{0}$. Since the depolarizing channel and the unitary channel commute, for noisy sensing, we let the unitary act noiselessly on a depolarized input state, which is given by 
\begin{equation}
    \tilde{\rho}_{0} = (1-q)\rho_{0} + q\frac{\mathbb{1}}{2},
\end{equation}
where $q$ is the depolarization strength. Noting that the eigenstates  of $\tilde{\rho}_{0}$ and $\rho_0$ are the same, Eq.~\eqref{eq:qfi} evaluates to
\begin{equation}
    \mathcal{F}(\tilde{\rho}_0) = (1-q)^2 \mathcal{F}(\rho_0).
\end{equation}
Note that the above expression is generically true for all input states $\rho_0$, meaning that the effect of noise is to simply scale the Fisher information by $(1-q)^2$, regardless of the input state. Therefore, the optimal state to use for noisy sensing is same as that for noiseless sensing.

\end{document}